\documentclass[prd,nofootinbib,preprintnumbers,floatfix]{revtex4}  
\usepackage[plainpages=false, colorlinks=true, anchorcolor=blue, linkcolor=blue, citecolor=blue, bookmarks=false]{hyperref}
\newcommand{\rthis}[1]{\textcolor{black}{#1}}
\usepackage{graphicx}

\begin{document}
\author{Pragna  \surname{Mamidipaka}$^{1}$}%
 \altaffiliation{ee20btech11026@iith.ac.in}
 \author{Shantanu \surname{Desai}$^{2}$ }%
 \altaffiliation{shntn05@gmail.com}
\affiliation{$^{1}$ Department of Electrical Engineering, Indian Institute of Technology, Hyderabad, Kandi, Telangana-502284  India}

\title{Do Pulsar and Fast Radio Burst dispersion measures obey Benford's law?}
\affiliation{$^{2}$ Department of Physics, Indian Institute of Technology, Hyderabad, Kandi, Telangana-502284  India}
\date{\today}

\begin{abstract}
We check if the first significant digit of the dispersion measure of pulsars and Fast Radio Bursts (using the CHIME catalog) is consistent with the Benford distribution.
We find a large disagreement with Benford's law  with $\chi^2$ close to 80 for 8 degrees of
freedom for both these aforementioned  datasets. This corresponds to a discrepancy of about 7$\sigma$. Therefore, we conclude that the dispersion measures of pulsars and FRBs do not obey Benford's law.
\end{abstract}
\maketitle

\section{Introduction}
Many naturally occurring distributions tend to adhere to a logarithmic distribution as predicted by Benford's law~\cite{Pinkham,Hill98}, which is sometimes often referred to as the significant digit law or the law of anomalous numbers.  This law states that given a distribution of numbers, the fraction of numbers with leading digit $d$ is given by~\cite{newcomb,benford}:
\begin{equation}
P(d) = \log(1+\frac{1}{d})
\label{eq:pd}
\end{equation}
This law has been widely applied to  an assortment of fields, including  biology~\cite{bio1,bio2}, finance~\cite{finance1,finance2}, geophysics~\cite{geo},  seismology~\cite{seis}, spectroscopy~\cite{whyman,bormashenko}, finance to detect banking frauds~\cite{finance2}, Nuclear and Particle Physics~\cite{buck,ni,Farkas,Dantuluri}, etc. This  law has been derived using a central-limit-like theorem for significant digits~\cite{Hillb}, as well as using Markov process~\cite{Burgos}.

 Benford's law has also been widely applied to a variety of astrophysical datasets, such as distances to stars and galaxies~\cite{astro2,astro3}, masses, orbital periods and semi-major axes of asteroids~\cite{Melita}, GAIA-2 parallaxes~\cite{Jong}, light curves of cataclysmic variables and other X-ray transient sources~\cite{Moret06}, etc.

 In this work, we check if the dispersion measures of pulsars  and Fast radio bursts (FRBs)  obey Benford's law. Pulsars are rotating neutron stars which emit pulsed emission in radio waves, and have been widely used as probes of nearly all branches of Physics and  Astronomy~\cite{handbook}.  FRBs are short-duration radio bursts located at extragalactic distances~\cite{FRBreview}. 
 Previously, Benford's law has also shown to be true for for a whole slew of other pulsar properties such as the time derivative of the barycentric period, first and second time derivative of rotation frequency, period derivative, spin down age, proper motions, spin-down luminosity, fluxes, transverse velocity~\cite{Shaopulsar}. However, this same work also showed that Benford's law does not hold true for barycentric period and barycentric rotation frequency.  Dispersion measure (DM) is the integrated free electron column density in the ionized interstellar medium. Precise measurements of DM  for millisecond pulsars is one of the main goals of various pulsar timing array experiments~\cite{KK,Noble}.

\section{Dataset and Analysis}
\subsection{Dataset}
For this work, we downloaded  the DM (in units of  $cm^{-3} pc$) for the radio pulsar population, along with the measurement uncertainties  from the ATNF online catalogue (version 1.67)~\cite{ATNF}~\footnote{ \url{http://www.atnf.csiro.au/research/pulsar/psrcat/}}. The dataset contains 3319 pulsars. We removed 122 pulsars for which no DM measurements were available.  We did this analysis for all the remaining 3197 pulsars with DM measurements  as well as a smaller sample of 3143 pulsars for which fractional errors in DM were less than 10\%.

For FRBs, we used  the DM measured from the  CHIME sample~\cite{CHIME}, which contains 536 FRBs in the same units as the pulsar population. This sample includes 474 one-off  events and 62 repeaters. None of the FRBs had fractional errors greater than 10\%. So there was no need to cull any objects from the FRB sample for our analysis.

\subsection{Analysis}
\begin{figure*}
    \centering
    \includegraphics[width=14cm,height=14cm]{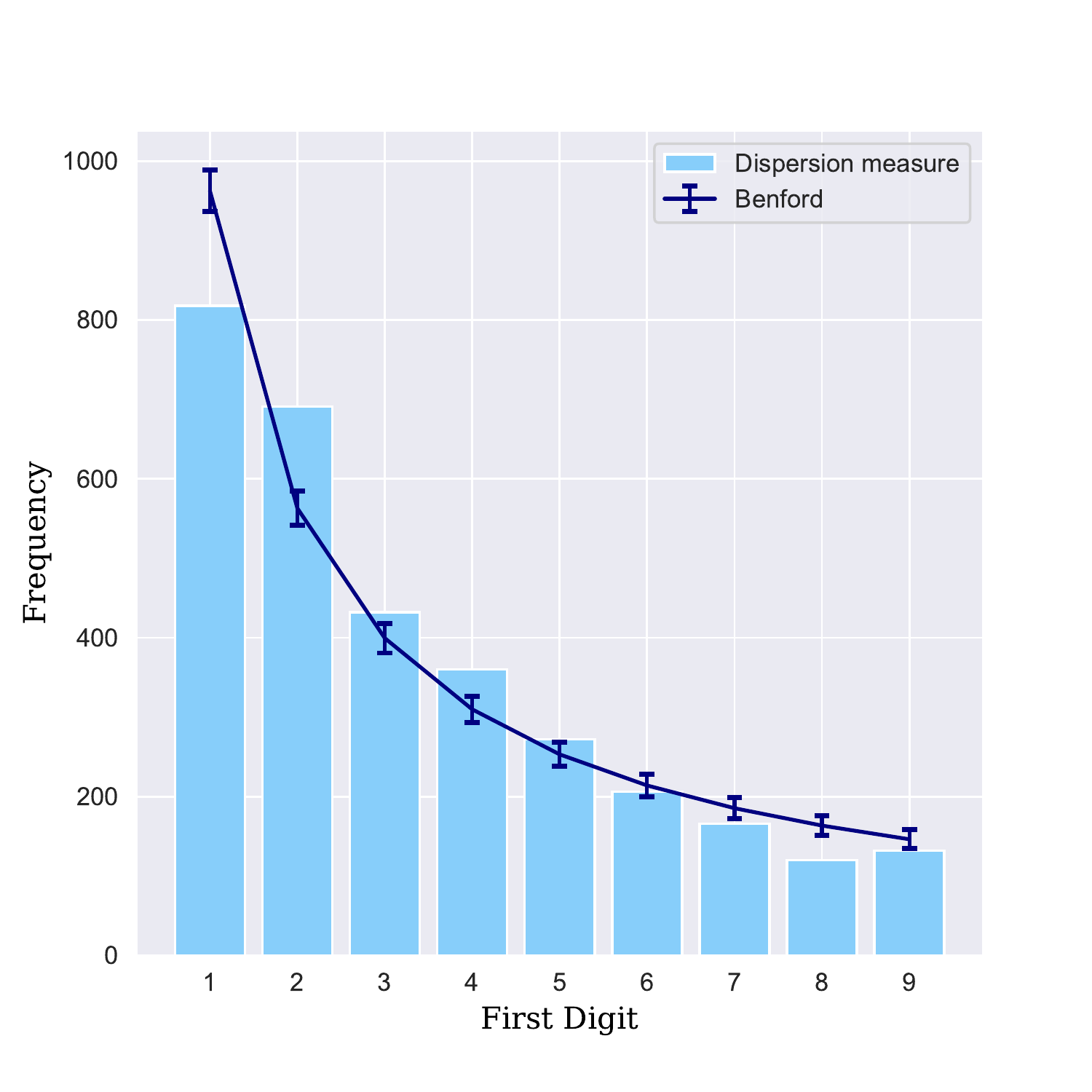}
    \caption{Distribution of the first digit for DM of 3197 pulsars from the ATNF catalog.}
    \label{fig1}
\end{figure*}

\begin{figure*}
    \centering
    \includegraphics[width=14cm,height=14cm]{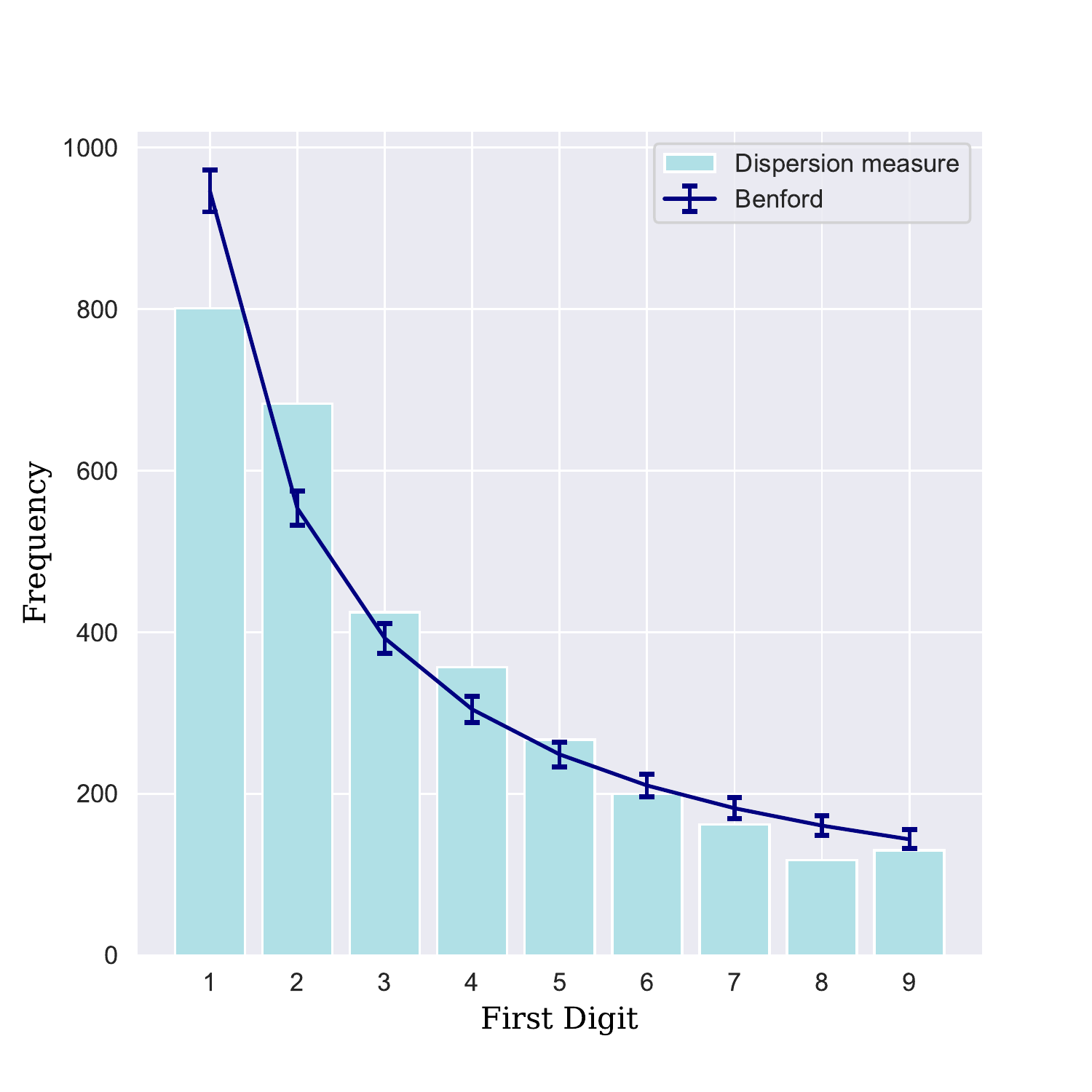}
    \caption{Distribution of the first digit for the DM of 3143 pulsars downloaded from the ATNF catalog, for which the fractional errors in DM were less than 10\%.}
    \label{fig2}
\end{figure*}

The first digit distribution for the complete pulsar sample can be found in Fig.~\ref{fig1}, whereas the same distribution after removing the data points in which the fractional errors are less than 10\% can be found in Fig.~\ref{fig2}. For both the figures, the expected Benford distribution along with associated binomial errors $\sqrt{N P(d)(1-P(d))}$ are also shown. By eye, we see that the Benford distribution is not obeyed although the relative rank of the frequency of leading digits agrees with Benford's law. Most of the discrepancy occurs for the first two digits which undershoot (1) and overshoot (2) the expected values respectively.

In order to quantify how well the pulsar dispersion measures adhere to Benford's law we use Pearson  $\chi^2$
\begin{equation}
\chi^{2} =\sum_{d=1}^{9} \frac{\left(N_{B}(d)-N_{O}(d)\right)^{2}} {N_{B}(d)} ,
\end{equation}
where  $N_{B}(d)$ indicates the expected count of occurrences according to Benford's law and $N_{O}(d)$ is the  observed number of occurrences for a single digit $d$. 
 For the full sample of pulsars, we obtain  a $\chi^{2}$ value of $78$ for 8 degrees of freedom. If we discard the measurements for which the fractional error in the dispersion measure is more than $10\%$, we obtain a Pearson $\chi^{2}$ value of $81$. Both these measurements  
correspond to $p$-values of $10^{-13}$ and $10^{-14}$, or in terms of $Z$-score , a discrepancy of 7.3$\sigma$ and 7.6$\sigma$, respectively,  using the prescription in~\cite{Cowan,Ganguly}.

\begin{figure*}
    \centering
    \includegraphics[width=14cm,height=14cm]{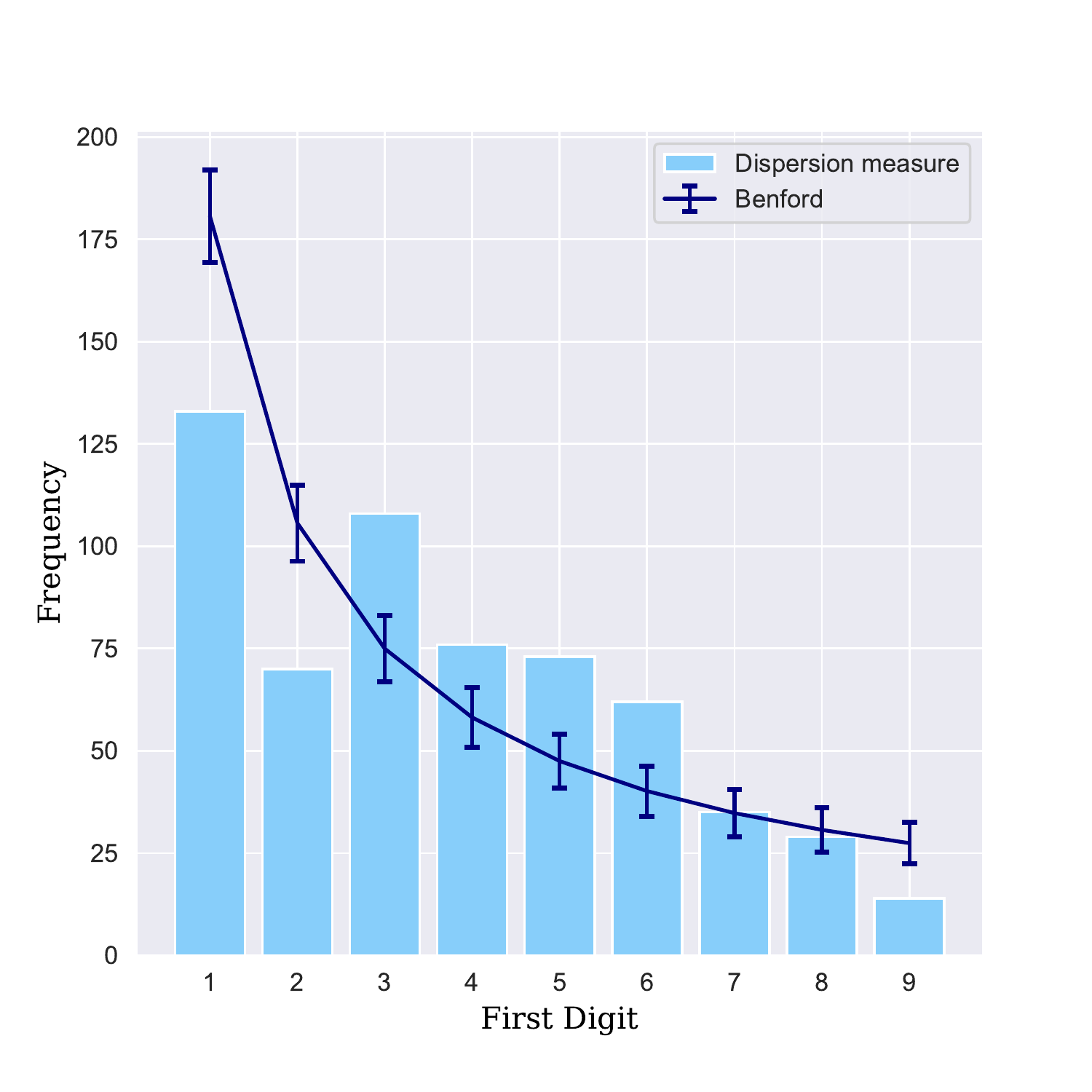}
    \caption{Distribution of first digit of the DM of all FRBs detected by CHIME~\cite{CHIME}}
    \label{figfrb}
\end{figure*}

The first digit distribution of FRBs can be found in Fig.~\ref{figfrb}. As we can see, the first two digits are much smaller than  the expectations due to Benford distribution.
The $\chi^2$ value we obtain is about 77 for 8 degree of freedom, corresponding to a $p$-value of $3 \times 10^{-13}$ or $7.2\sigma$ discrepancy.

A tabular summary of all our results can be found in Table~\ref{tab:results}. We conclude that both pulsar and FRB DMs do not adhere to the Benford distribution.

\begin{table}
\begin{center}
\begin{tabular}[t]{|l ||c||c|| c|}
\hline\hline
\textbf{Dataset} &  $\chi^{2}/dof$  & $p$-value & \textbf{Discrepancy Significance} \\ 
\hline
Pulsar (Full)		&	78/8   &  $10^{-13}$  &   7.3$\sigma$ \\
Pulsar ($\sigma_{DM}/DM <$ 10\%) & 81/8 & $10^{-14}$ & 7.6$\sigma$ \\
FRB & 77/8 & $3 \times 10^{-13}$ & 7.2$\sigma$  \\
\hline\hline
\end{tabular}
\end{center}
\caption{Summary of our results on Benford analysis of first digit of DM of pulsars and FRBs. As we can see, none of the datasets agree with Benford's law.}
\label{tab:results}
\end{table}

\section{Conclusions}
In this work the first significant digit of the DM for pulsars and FRBs is examined for adherence to Benford's distribution. For pulsars we considered the full dataset for which DM measurements were available, as well as those for which the fractional errors in DM were less than 10\%. For FRBs, we used the dataset from CHIME catalog.

The distribution of first digit of DM for the aforementioned datasets can be found in Figs.~\ref{fig1}, ~\ref{fig2}, and ~\ref{figfrb}. None of the three datasets agree with Benford distribution. A tabular summary of our results can be found in  Table~\ref{tab:results}.
All the datasets show a disagreement with Benford's law with significance between 7.2-7.6$\sigma$. 
Therefore, we conclude that the DM of pulsars and FRBs do not agree with Benford's law, even though some other properties of pulsars have been previously shown to adhere to Benford's law~\cite{Shaopulsar}.

\rthis{The main reason for this, is that  given  the physical location of pulsars/FRBs, there is not enough dynamic range  in the data for them to adequately obey the  Benford  distribution.}

\rthis{For pulsars, there is a dearth (only 6\%) of objects with DM $< 20 cm^{-3} pc$. Majority of the pulsars (35\%) have  DMs between 20 and 100 $cm^{-3} pc$ and only about 20\% of them have DMs of between 100-200 $cm^{-3} pc$. This is  due to the fact that the  detection of pulsars is limited by the sensitivity of radio telescopes, and the DM distribution of pulsars is very much direction dependent. Therefore, the number of occurrences  of the number one  in the first digit of the DM is suppressed, whereas the same is enhanced for the  number two compared to the expectations from Benford's law.}

\rthis{Similarly, since FRBs  are located at cosmological distances, only a small fraction of  FRBs (8\%) have DM $< 200 cm^{-3} pc$, since there are very few Galaxies nearby. Furthermore, there are very few FRBs (15\%) with DM $> 1000 cm^{-3} pc$, since radio telescopes are not sensitive to very high redshift FRBs.  Therefore, the full dynamic range of the DMs for FRBs is  only about a  factor of 20-30, and the number of occurrences of the number one in the  first digit of DM is suppressed compared to Benford's law.}

\rthis{Therefore for the above reasons, the DMs of pulsars and FRBs do not obey Benford's law, and therefore our results are not unexpected or a surprise.}

\begin{acknowledgements}
\rthis{We are grateful to the anonymous referee for useful  feedback on our manuscript. }
\end{acknowledgements}

\bibliography{main}
\end{document}